 \documentstyle[12pt]{article}

\renewcommand{\baselinestretch}{1.05}

\begin{document}

\def\footnoterule{\hrule width \hsize}
\skip\footins = 18pt
\footskip     = 18pt
\footnotesep  = 15pt

\pagestyle{empty}

\begin{titlepage}

\vspace*{3cm}
\begin{center}
{\Large
FLAT DIRECTIONS AND BARYOGENESIS \\[0.6ex]
IN SUPERSYMMETRIC THEORIES}\\
\vspace*{36pt}
{\large Lisa Randall%
\footnote{\baselineskip=13pt
NSF Young Investigator Award, Alfred P.~Sloan Foundation Fellowship, DOE
Outstanding Junior Investigator Award.
This work is supported in part
by funds provided by
the U.S.~Department of Energy (D.O.E.)
under contract \#DE-FC02-94ER40818
and by N.S.F. grant PHY89-04035.}%
\renewcommand{\thefootnote}{\fnsymbol{footnote}}%
\setcounter{footnote}{0}%
\footnote{Based on work done in collaboration with Michael Dine and
Scott Thomas.}}
\\
{\sl Center for Theoretical Physics, Laboratory for Nuclear Science \\
and Department of Physics, Massachusetts Institute of Technology \\
77 Massachusetts Avenue, Cambridge, MA ~02139-4307 \\
E-mail: {\tt lisa@ctptop.mit.edu} }\\

\def\footstrut{\baselineskip 13pt}

\vspace{2cm}
\begin{abstract}
 Flat directions are a generic feature of supersymmetric theories.
They are of cosmological interest because they can lead to coherent
production of scalars.
In the early universe such flat directions  could be
dangerous due to the potentially  large energy density  and the late decay of
the associated
scalars when they have only $1/M_p$ couplings  (Polonyi
problem). On the other hand, flat directions among
the standard model fields can carry baryon
number and lead to a possible mechanism for baryogenesis
(Affleck Dine baryogenesis).   When considering the cosmological
consequences of the flat directions, it is important to take into account
the soft potential with curvature of order the Hubble constant
due to supersymmetry breaking in the early universe.
In this talk, we
discuss flat directions, their potential cosmological implications
focusing on Affleck-Dine baryogenesis,
and how the standard picture  of their
evolution must be modified in the presence of  the large supersymmetry breaking
in the early universe.
\end{abstract}

\end{center}

\vspace*{1.0cm}

\centerline{\it to appear in the Proceedings of the XXXth Rencontres
de Moriond}
\centerline{\it ``Electroweak Interactions and Unified Theories''}
\centerline{\it Les Arcs, France, March 1995}

\vspace*{1.0cm}

\end{titlepage}

\def\begs{\begin{slide}{}}
\def\ends{\end{slide}}
\def\begi{\begin{itemize}}
\def\endi{\end{itemize}}
\def\beq{\begin{equation}}
\def\eeq{\end{equation}}
\def\begsm{\begin{small}}
\def\endsm{\end{small}}
\def\mgravitino{m_{3/2}}
\def\mphi{m_\phi}
\def\nphi{n_\phi}

\renewcommand{\baselinestretch}{1.5}

Flat directions are a common feature of supersymmetric theories
and have many important implications. Here we will focus on
some of the cosmological aspects of the existence of flat directions
which can lead to the coherent production of a scalar condensate.
Before we proceed however, we address the question of why supersymmetric
cosmology is of interest. One well known reason is that
one can  potentially constrain particle physics models.
Chief among the requisite constraints are  by  that
the universe is not overclosed and that nucleosynthesis can
proceed successfully. This latter constraint gives rise to the
cosmological moduli (Polonyi) problem \cite{1,2,3} for example.
For a more extensive  discussion of the potential problems, see ref.  \cite{11}

A related reason for studying supersymmetric
cosmology is the difficulty
of experimentally probing the high scales associated with supersymmetric
physics.  Cosmology offers an alternative probe to high
physics scales like that associated
with supersymmetry breaking, flavor, or GUTS. In particular, operators which
are suppressed
by high mass scales might nonetheless be relevant at early times
when fields take large expectation values.   Unfortunately,
this also means  that if a mechanism  of
baryogenesis exists which exploits CP violating
high dimension operators, this nonstandard model
CP violation is experimentally inaccessible, so that there is not
necessarily a connection between CP violation which will be explored
in a laboratory and that which was important for baryon number creation.

A third reason is that we know there is a baryon asymmetry in the universe
and it is worthwhile to understand its origin. GUT scale baryogenesis
must contend with $B+L$ violation at late times and a relatively
low reheat temperature after inflation. Weak scale baryogenesis
is a nice possibility, but there are many difficult questions regarding
the nature of the phase transition.  It is useful to study in detail
alternative mechanisms for baryogenesis. The Affleck Dine (AD) mechanism
\cite{4}
in particular is a beautiful way to utilize the flat directions which
are in any case present in the supersymmetric standard model
for the creation of baryons.  Although there
this mechanism has been studied  in the past, we show that the more likely
picture
of how the AD mechanism works is substantially different from
what has been studied, and yields qualitatively and quantitatively
different conclusions.

In this talk, we will first review the flat directions of supersymmetric
theories,
and the old picture for coherent production in the early universe.
We will argue that this picture is modified because of supersymmetry
breaking in the early universe, which effectively
generates a soft superysymmetry
breaking scale of order $H$, where $H$ is the Hubble constant.
When the Hubble constant is bigger than $m_{3/2}$ which
is of order of the weak scale, the supersymmetry  breaking
potential
determined by $H$ scale terms is dominant, and changes significantly
the picture of the evolution of the fields at early times.

We will then  discuss two implications of this revised picture
of the early universe field evolution.  We will show that it
changes the standard scenario for the Polonyi problem, and potentially
presents a solution.  The major focus of this talk however will
be the implications for baryogenesis through the Affleck-Dine \cite{7}
mechanism.  We will discuss the evolution of the Affleck-Dine
field in the presence of the Hubble constant scale potential
and also include higher dimension operators which can generate
a potential for the ``flat" directions, even in the supersymmetric limit.
We will see that we have a relatively simple and predictive scenario
for baryogenesis. We find that depending on the identity of the AD
field, one can naturally obtain $n_b/s\ge 10^{-10}$. This is
in contrast to the previous picture, according to which additional
entropy deposition in the late universe was required.

Flat directions are   peculiar to supersymmetric theories.
They correspond to fields with no classical  potential.  In the absence of
supersymmetry,
they would be highly unnatural.   However, supersymmetric nonrenormalization
theorems protect massless fields to keep them massless, even
with radiative corrections.  There
are several contexts in which flat directions  are relevant.  One is  the
moduli space (of string theory for example) which  parameterizes
a large vacuum degeneracy of physically inequivalent theories not related
by a symmetry.  These flat directions  might have
no perturbative superpotential couplings.  However, even  if they do
   couple in the superpotential,
one can  often find combinations of fields for which there is no potential
due to accidental degeneracies.
This happens even in the supersymmetric standard model, where
there are a large number of such directions.

A simple example of such a flat direction is when the $H_u$ and $L$
fields have equal expectation values so that the $D$ term contribution
to the potential is cancelled.    Another example
of a flat direction of the supersymmetric standard model is $Q_1 L_1
\bar{d}_2$, where the numerical index
labels generation number.
 There are many such flat directions for which both the $D$
and $F$ type contributions to the potential vanish.

It is not true however that the potential for flat
directions  vanishes identically. There
are two ways in which one expects the flat directions to be lifted. One
possibility is nonrenormalizable operators in the superpotential.
These are not necessarily present, but in many cases, if they
are consistent with all existing symmetries, one would expect
such operators to occur, suppressed by a high dimension scale
which might be $M_p$ or $M_G$ or some other high scale of the theory.
These operators turn out to be very important to the AD mechanism
because they are the source of $B$ (or $L$) and CP violation.

The other source of the potential for the flat direction fields is
soft supersymmetry breaking. In  the absence of additional
symmetry, one would expect all fields to get a mass of order $m_{3/2}$
when supersymmetry is broken. Of course, in the early universe,
when $m_{3/2}\ll H$, this is negligible.

  Why are these flat direction fields of cosmological relevance?
Assume a flat direction field has ``no" potential, so the initial
field value is undertermined. In this case, one would expect
some random initial displacement of the field from its zero temperature
minimum.  Consider now the classical evolution of the zero mode,
 \beq
\ddot \phi+3H\dot{\phi}+V'(\phi)=0
\eeq
 where $H$ is the damping term in an expanding background.
 It is clear that if
$H^2\gg V''$ we are in the  overdamped case and$\phi\approx$ constant,
whereas if
 $H^2\ll V''$, the field is  underdamped and $\phi$ evolves to minimum
and oscillates freely.
  These oscillations imply a coherent production of nonrelativistic particles
 This coherent production of scalar particles in the early universe is generic
in supersymmetric theories with flat directions.

Now what are the implications of these particles?
Clearly it will depend on the couplings and quantum numbers
of the field. If there are  only gravitational strength ($1/M_p$) interactions,
one finds the
 Polonyi problem. The standard
 statement of the  problem is as follows.
 Suppose the $\phi$ field has only $1/M_p$ couplings and  $m_\phi\approx
O(m_{3/2})$.
Suppose also that the initial field value  $\phi_0\sim M_p$. Then the field
is frozen
  until  $H\sim m_{3/2}$, after which it
begins to oscillate freely. At this point,  $\rho_\phi\approx m^2
M_p^2\Rightarrow \rho\approx \rho_{universe}$.
Subsequently, $\rho_\phi$ dominates the energy density of the
universe. When $H\approx \Gamma\approx m^3/M_p^2$, $\phi$ decays
and the associated reheat temperature
 $T_R\approx \sqrt{\Gamma M_p}\approx 10{\rm keV}$. This is too low
for successful nucleosynthesis.    Moreover, even  if the density of the
condensate
is somewhat lower, late decays of the condensate would destroy
the nuclei which have already been created, so the problem is
even more severe than it naively appears.

An alternative scenario for the coupling of the flat direction is that
the field is a flat direction of the standard model carrying net $B-L$.
 In general, the AD mechanism requires that $B-L$ is violated
at large field value, but conserved for small field value.
It is assumed that the field at early times is displaced from
its true minimum. Eventually, the field is driven towards
the origin through the equations of motion. The mechanism
works when baryon number violation is important as the field moves
in towards the origin, so that baryon number is
stored in the coherent condensate. Once the field is oscillating about the
minimum,
the baryon number violating operators are no longer significant and baryon
number
is conserved. Notice that the three conditions for baryogenesis \cite{6}
can be satisfied. There is  CP violation through the phase difference between
the initial phase of the field and that of the baryon number
violating operator. There is $B$ violation by assumption (although
we have not yet specified the source). Finally,
the large initial value for the field $\phi$ is a nonequilibrium situation.

 However, at this point, we clearly do not have the whole story. We would
like to better understand what are the initial field values, what
provides the baryon number violation (for the AD mechanism),
and whether the flat directions really are as flat as we have been assuming.
In the rest of the talk, I argue that the picture we have been presenting
is not the whole story.   The most significant aspect which has been
omitted is that supersymmetry is necessarily broken in the early
universe. Furthermore the scale which acts as the soft supersymmetry
breaking parameter is $H$, not $m_{3/2}$ \cite{7,8,9}.  This means that
the picture of the fields as being ``frozen" at early times is not correct.
In the remainder of the talk, we explore the consequences.

We will briefly discuss the implication for the Polonyi problem.
As for the AD mechanism, we will find that one can incorporate
the $B$ violation in higher dimension operators in the superpotential.
We will find that the AD mechanism does not always work; what
is required is that the effective mass squared in the early universe is
negative
in order to drive the AD field { classically} to large field value.  However,
at this point, all that is needed in order to derived $n_b/s$ will be the
dimension of the superpotential operator which stabilizes the potential
and the reheat temperature after inflation. The picture of the evolution
of the AD field is significantly altered, but a very elegant scenario emerges.

First let us understand why supersymmetry is necessarily
broken in the early universe.  The Hubble constant
$H$ is related to the matter energy density  $\rho$ by $H^2=\rho/M_p^2$.
Therefore, an expanding universe
 implies a finite positive energy density in the early universe,
implying supersymmetry is broken.  In the case the
energy is carried by radiation, supersymmetry is broken by  the different
thermal occupation
numbers of bosons and fermions.  However this is well known and
is not the effect which is of interest to us, since by the time
the universe is radiation dominated after inflation, the zero
temperature supersymmetry breaking is dominant. The case
we will be most interested in is during and subsequent to inflation,
when the Hubble constant is very large compared to $m_{3/2}$.  Notice
that this is in accord with the well known result that supersymmetry
is broken in deSitter space.

 We now consider the potential which is generated for the flat direction
field.
 Suppose $\rho=F_I^\dagger F_I$ and $K\supset \phi^\dagger \phi I^\dagger
I/M_p^2$.
Then $m_\phi^2\approx F_I^\dagger  F_I/M_p^2\approx H_I^2$.
In fact, more generally if  $\langle O \rangle =\rho$ and there is
an  operator in the potential of the form
$O \phi^\dagger \phi/M_p^2$,
then $m_\phi^2\approx \rho/M_p^2\approx H^2$.   So we see there
is generally a soft supersymmetry breaking mass in the early universe
of order $H$. This Hubble scale mass is essential to the evolution
when $H>m_{3/2}$.

One can ask whether this mass is necessarily present.  The answer
is yes, unless there is fine tuning.  The point is that even with {\it minimal}
Kahler potential, such a mass term occurs.  Recall
the supersymmetric potential takes the form
  \beq
V=e^K\left(\left(W_i+{K_i W\over M_p^2}\right)K^{i\bar{j}}\left(W_j+{K_j W\over
M_p^2}\right)^*-3{W^*W \over M_p^2}\right)+{1 \over 2} g^2D^aD^{a}
\eeq
Suppose there is minimal $K=\phi^\dagger \phi$. Then one can see that
 there is a  mass of order $H$  when the potential energy
is finite and positive, and with the
 minimal Kahler potential the Hubble scale mass is positive.

With a nonminimal Kahler potential (or $I\approx M_p$) the mass formula is
more complicated.
 For example, if
 \ $ K\supset \phi^\dagger \phi I^\dagger I/M_p^2$, then
$K_{I\bar{I}}=\phi^\dagger \phi/M_p^2\Rightarrow m^2\subset -F_I^\dagger
F_{\bar{I}}/M_p^2\approx -H^2$.
Notice that a nonminimal $K$ is to be expected.
In fact,  such higher dimension operators are necessarily present
in the Kahler potential as counterterms for the running of the
soft masses \cite{10}. And as we will see later,
the standard model running of the soft mass might be
adequate to give negative mass squared to $LH_u$, the preferred AD field
(as we will argue).

So we conclude there is certainly a mass of order $H$
but its sign  and magnitude is not determined.

We now consider the field evolution in the presence of the potential
due to supersymmetry breaking in the early universe.  The first
observation is that generically, the fields are not frozen; that
is they evolve to a local minimum as they are
not overdamped,  since $m\approx H$.   For example,
with minimal Kahler potential, the field will roll to the origin.

What does this imply in terms of the Polonyi problem?  It should be
clear that the ``initial" field value is not random; the field evolves
in the effective potential to  a nearby local minimum. This might
be a concrete realization of the problem, since in general, the minima
in the early universe and today do not coincide. However it also
suggests a solution to the problem in the cases where these minima
are the same. This might be true for example when the minimum
is a point of enhanced symmetry \cite{7,8}.  It can also arise as a consequence
of a factorization of the Kahler potential so that minima of $K$
are local minima of the potential \cite{11}.

It should be noted however that an exception to this scenario is
the case where additional symmetries protect the mass of the
flat direction field, eg a pseudogoldstone boson. Then the associated
symmetry breaking parameter is necessary to generate a mass,
so that the mass will generically be suppressed relative to $H$.
In this case, one expects the field to always be overdamped.

What are the implications of this Hubble scale mass for the Affleck
Dine mechanism? First suppose the mass squared was positive.
Then the field will be classically driven to the origin, in which
case the amplitude of the AD field vanishes, and there is no baryon number
creation! Previous authors suggested that quantum, not classical
effects, drive the field away from the origin. But this will not
work because when $m\approx H$, the coherence length is too small
and baryon number would average to zero over the observable universe.

However, we have argued that the Hubble scale mass squared
is not necessarily positive.  In fact, we will now show that
the negative mass squared scenario works perfectly. The AD field
is driven classically to a large field value. When $H\approx m_{3/2}$,
the low energy supersymmetry breaking gives a comparable
contribution to the potential as the $H$ dependent terms.  About
this time, the mass passes through zero to become positive, and
the AD field rolls towards the true minimum at the origin.  What
makes this whole scenario work so efficiently is that the baryon number
violating operators turn out to also be comparable to other
terms in the potential at this time, so that baryon violation is essentially
maximal.  The condensate stores baryon number, and $n_b/s$
is determined by the relative fraction of the AD field at this time.

Let us examine in more detail the salient features of
the evolution. It will be useful
to divide our analysis into three periods; during inflation, the
post inflation-inflaton matter dominated era when $m_{3/2}\ll H$,
and the post inflation-inflaton matter dominated era when $m_{3/2}\approx H$.
In our analysis, we assume the AD potential arises from
\begi
\item Nonrenormalizable terms in the superpotential
\item Soft masses (and possibly soft $A$ type terms) of order $H$
due to supersymmetry breaking in the early universe
\item Zero temperature supersymmetry breaking parameters
whose scale is determined by $m_{3/2}$. These are negligible
in the early stages of the evolution, but important when baryon
number is established and subsequently, that is, for $H\le m_{3/2}$.
\endi

Now let us consider the field evolution during the three
epochs outlined above. First consider the potential during inflation.
\beq
V(\phi) = -cH_I^2 |\phi|^2 +
\left( {a\lambda H_I \phi^n \over n M^{n-3}} + ~h.c.~ \right) +
|\lambda|^2{|\phi|^{2n-2} \over M^{2n-6}}
\label{veff}
\eeq
where $c$ and $a$ are constants of ${\cal O}(1)$, and
$M$ is some large mass scale such as the GUT or Planck
scale.
For $H_I \gg \mgravitino$ soft terms arising from the hidden
sector are of negligible importance.
 The minimum of the potential (\ref{veff}), is given by
\beq
|\phi_0| = \left( { \beta H_I M^{n-3} \over  \lambda }
\right)^{1 \over n-2}
\label{phimin}
\eeq
where $\beta$ is a numerical constant which depends on
$a$, $c$, and $n$.
Notice that $\phi_0$ is parameterically between
$H_I$ and $M$.

\begi
\item During inflation, the AD field evolves exponentially
to the minimum of the potential, determined by the induced
negative mass squared and nonrenormalizable term in the superpotential.
This process may be thought of
as establishing ``initial conditions" for the subsequent
evolution of the field.  In the presence of a supersymmetry breaking
``A" type term, the phase will also roll to its minimum. Otherwise
a random value for the phase is taken. Either way, it is the difference
between this phase and the real minimum for the phase which
establishes baryon number.

It is easy to determine that the field rolls efficiently to its minimum.
Suppose it started far away since the field is rapidly oscillating
with a slowly decreasing envelope.  The time rate of change of the energy in
$\phi$ can be found
from the equations of motion,
\beq
{dE \over dt}=\dot{\phi}{d \over d\phi}(T+V)=-3H_I{\dot{\phi}}^2
 = -6H_I(E-V)
\eeq
where $T$ is the kinetic energy.
Using the expression for $V(\phi)$ for large $\phi$ and
averaging over a period gives
$\dot\phi_m \simeq -6H_I/(2n-1) \phi_m$.
We therefore conclude that in the large $\phi$ regime,
$\phi$ decreases exponentially towards smaller values,
\beq
\phi_m \simeq e^{-6H_It/(2n-1)}\phi_i
\eeq
where $\phi_i$ is the initial value of the field with
espect to the origin.
Thus after just a few $e$-foldings $\phi$ is near a minima.  Once
near a minimum, the field evolves like a damped harmonic
oscillator.

\item  After inflation the universe enters a matter era dominated by the
coherent oscillations of the inflaton. The minimum of the potential
is time dependent (as it is tied to the instantaneous value of the
Hubble parameter). The AD field oscillates
near this time dependent minimum with decreasing
amplitude.

During a matter era the Hubble constant is related to the expansion
time by $H=\frac{2}{3}t$.
The equation of motion for $\phi$ is then
\beq
\ddot \phi + \frac{2}{t} \dot \phi + V^{\prime}(\phi) =0
\label{phiteq}
\eeq
where $V(\phi)$ is still given by (\ref{veff}),
though the dimensionless
constants $c$ and $a$ may be different, and $H$ is now
time dependent.

We can obtain greater insight into the solutions of (\ref{phiteq}),
and also obtain a form more suitable for numerical study by
making changes of variables. We define
 $$
z=\log t.
$$
It is also useful to define  the dimensionless field $\chi$
by scaling with respect to the instantaneous minimum
of the effective potential.
$$
\phi=\chi \phi_0(t) = \chi
 \left( \frac{ \beta}{ \lambda} M^{n-3} e^{-z}
   \right)^{\frac{1}{n-2}}
$$
where $\beta=\sqrt{c^{\prime}/(n-1)}$ for $a=0$, and
$c^{\prime}= \frac{4}{9} c$.
The equation of motion in these rescaled variables is then
\beq
\ddot\chi +
\left( \frac{n-4}{n-2} \right) \dot\chi -
\left[ c^{\prime} + \frac{n-3}{ (n-2)^2 } \right] \chi +
    c^{\prime} \chi^{2n-3} = 0
\label{rescaledeom}
\eeq
The rescaled problem is so simple because the
effective mass term, Hubble damping term, and acceleration
term are all homogeneous in $z$.

The qualitative behavior of the solution is now much more apparent.
We see that we have eliminated all large and small parameters
from the differential equation.  Unless the damping term is negative,
we expect the field to track the minimum of the true potential.  Notice
that the effective potential for the rescaled variable has a new contribution
to the effective mass but the AD  field  is of the same order of magnitude as
if it were at the minimum of the potential.

We now see that for $n>4$, the field is driven towards the minimum, while
for $n<4$, it would be driven away.  This latter case would however correspond
to a field which was not flat, so it is not of interest to us.  The case $n=4$
is
interesting in that the rescaled field is not driven closer to the minimum than
its  iniital value, although the true field is due to the scaling.

 In any case,
it should be clear that in all cases of interest, the field amplitude
essentially
follows the effective minimum determined by balancing the time dependent
(Hubble
dependent) mass term and the nonrenormalizable term in the superpotential.
Because
the mass is decreasing with time, the field amplitude decreases with time.

 \item When $H \sim \mgravitino$ the soft potential arising from
hidden sector supersymmetry breaking becomes important and the sign
of the mass squared becomes positive.
At this time, the $B$-violating $A$ term arising from the hidden
sector is of comparable importance to the mass term,
thereby imparting a substantial baryon number to the condensate.
The fractional baryon number carried by the condensate is near
maximal,
more or less independent of the details of the flat direction.  Subsequent
to this time, the baryon number violating operators are negligible
so the baryon number (in a comoving volume) is constant.

 The potential takes the form
\beq
V(\phi) = m_{\phi}^2 |\phi|^2
 - { \frac{c^{\prime}}{t^2}} |\phi|^2
+ \left( {(A m_{3/2} + aH) \lambda \phi^n \over n M^{n-3}} ~+h.c.~ \right)
 + |\lambda|^2 { |\phi|^{2n-2} \over M^{2n-6}}
\label{vtot}
\eeq
where $m_{\phi} \sim \mgravitino$.
At early times the field tracks near the
time dependent minimum as discussed in the last section.
Therefore when $H \sim \mgravitino$ all the terms in (\ref{vtot})
have comparable magnitudes.
Since the soft terms have magnitudes fixed by $\mgravitino$
the field is no longer near critically damped, but becomes
underdamped as $H$ decreases beyond $\mgravitino$.
In addition the $\mphi^2 |\phi|^2$ term comes to dominate
the $-cH^2 \phi^2$ term as $H$ decreases.
The field therefore begins to oscillate freely
about $\phi=0$
when $H \sim \mgravitino$, with ``initial'' condition given
by $\phi_0(t)$ (eq. (\ref{phimin})) with $t \sim \mgravitino^{-1}$.

Crucial for the generation of a baryon asymmetry are the $B$ violating
$A$ terms in (\ref{vtot}).
However,  when $H \sim \mgravitino$ all the
terms have comparable magnitude, including the $A$ terms.
Since $V_B \sim V_{\not B}$ when the field begins to oscillate freely
a large fractional baryon number is generated in the ``initial''
motion of the field when $m^2$ becomes positive.
Notice that in this negative mass squared scenario $n_b / \nphi$
is roughly {\it independent} of $\lambda / M$.
This is because the value of the field is determined precisely
by a balance of (negative) soft mass squared term and
nonrenormalizable supersymmetric term.
That the $B$ violating $A$ term also has the same magnitude
follows from supersymmetry since its magnitude is the root mean square
of the soft mass term and nonrenormalizable
supersymmetric term.
In this scenario there is no need for ad hoc assumptions about
the initial value of the field when it begins to oscillate
freely.
The expectation that $n_b / \nphi \sim {\cal O}(1)$ falls
out naturally.

We also did  numerical simulations to confirm
the above conclusions.
Again it is useful to rescale variables. The field is rescaled as
$$
\phi \rightarrow\left( \frac{ \mgravitino M^{n-3} }{ \lambda }
   \right)^{\frac{1}{n-2}}\phi
$$
{}From the arguments above , up to a numerical constant
of order unity this is just the value of the field when
$H \sim \mgravitino$.
All other mass scales and time are rescaled with
respect to $\mgravitino$.
The equation of motion (\ref{vtot}) with $a=0$ and
$\theta_A + \theta_{\lambda} =0$
is then
\beq
\ddot{\phi} + \frac{2}{t} \dot{\phi}
 + \left( \mphi^2 - \frac{c^{\prime}}{t^2} \right) \phi
 + A \left( \phi^* \right)^{n-1}
 + (n-1) \left( \phi^* \phi \right)^{n-2} \phi
   = 0
\eeq
Notice that the independence from $\lambda / M^{n-3}$ is
manifest in this form.
The equation of motion for the real and imaginary parts
(appropriate for numerical integration) are
\begin{eqnarray}
 & &   \ddot\phi_R
   + \frac{2}{t} \dot{\phi}_R
   + \left( \mphi^2 - \frac{c^{\prime}}{t^2} \right)  \phi_R
   + A |\phi|^{n-1} \cos \left( (n-1) \theta \right)
   + (n-1) |\phi|^{2n-4} \phi_R
   = 0
     \nonumber \\
 & &   \ddot\phi_I
   + \frac{2}{t} \dot{\phi}_I
   + \left( \mphi^2 - \frac{c^{\prime}}{t^2} \right) \phi_I
   - A |\phi|^{n-1} \sin \left( (n-1) \theta \right)
   + (n-1) |\phi|^{2n-4} \phi_I
   = 0
    \nonumber \\
\label{rieom}
\end{eqnarray}
where $\phi = \phi_R + i \phi_I$, and
$\theta = {\rm Arg}~\phi$.
It is straightforward to integrate   these equations forward,
assuming the field begins near the minimum at a time
when the mass is still negative.
  When $t \sim 1$ ($H \sim \mgravitino$) the field feels a
``torque'' from the $A$ term, and spirals inward in the
harmonic potential.
The nonzero $\dot{\theta}$ in the trajectory gives rise to
the baryon number. At late time,  $n_b / n_{\phi}$ asymptotes to a constant
value.

At very late stages of the evolution when $H \ll m_{3/2}$,
the only potential term which is relevant in
(\ref{vtot}) is the soft mass term $\mphi^2 |\phi|^2$
which is of course $B$ conserving.
The baryon number created during the epoch $H \sim \mgravitino$
is therefore conserved by
the classical evolution of $\phi$ for $H \ll \mgravitino$.

This establishes that $n_b/n_0\approx 1$. What is important
however is $n_b/s$. This is determined  by both the baryon fraction
of the condensate and also by the  fractional density of matter carried
by the condensate.

\item The inflaton  decays when $H < \mgravitino$ (consistent
with the gravitino bound on the reheat temperature). The condensate
will decay soon afterwards through scattering with the thermal
bath (but see \cite{7} for some caveats).
The baryon to entropy ratio subsequent to inflation is determined from
\beq
{n_b \over s} \approx
{n_b\over n_\phi}{T_R \over m_\phi}{\rho_\phi \over \rho_I}
\eeq
where $n_b$ and $n_{\phi}$ are baryon
and AD field number densities, $T_R$ is the reheat temperature,
$m_\phi \sim \mgravitino $ is the low energy mass for the AD field,
and $\rho_\phi$ and $\rho_I$ are the AD field
and inflaton mass densities (both at the time of inflaton decay).
\beq
{\rho_{\phi} \over \rho_I} \approx \left (
{ \mgravitino M^{n-3} \over \lambda M_p^{n-2} }
   \right )^{2/(n-2)}.
\eeq
For $n=4$,
$ \rho_{\phi} / \rho_I \sim 10^{-16} (M / \lambda M_p)$,
while for $n=6$,
$ \rho_{\phi} / \rho_I \sim 10^{-8} (M^3 / \lambda M_p^3)^{1/2}$,
Notice for smaller $(\lambda/M^{n-3})$
the direction is effectively flatter, and $\phi_0$ and $\rho_{\phi}$
are larger.
A greater total energy is therefore stored in the oscillating condensate
for smaller $\lambda$ or larger $n$.
\endi
Notice that for $n=4$ and high reheat temperature (motivated by naturalness
arguments), $n_b/s$ turns out to be just about right. It is worth
noting that all flat directions can be lifted by operators with $n\le 6$,
so that it is conceivable that any direction can give the correct $n_b/s$
without
additional entropy dump, although $n=6$ requires a low reheat temperature,
of order the weak scale.

The $LH_u$ direction appears to be especially promising for natural
production of the correct baryon to entropy ratio.
The only directions which carry $B-L$ and can be lifted at $n=4$ in
the standard model are the $LH_u$ directions.
The nonrenormalizable operator is then
\beq
W = \frac{\lambda}{M} (LH_u)^2
\label{LHLH}
\eeq
At low energies this is the operator which gives rise to
neutrino masses.
For baryogenesis along the $LH_u$ direction in this scenario,
$n_b/s$ can therefore be related to the
{ lightest} neutrino mass since the field moves out furthest
along the eigenvector of $L_i L_j$ corresponding to the
smallest eigenvalue of the neutrino mass matrix.
\beq
\frac{n_b}{s} \sim 10^{-10} \left( \frac{T_R}{10^9~{\rm GeV}} \right)
 \left( \frac{10^{-5}~{\rm eV}}{m_{\nu}} \right)
\eeq
If $T_R < 10^9$ GeV in order to satisfy the gravitino
bound \cite{12}, then  at least one must neutrino be
lighter than roughly $10^{-5}$ eV.

Another reason this field is so interesting as a candidate for the AD field
is that even with {\it minimal} Kahler potential, the scenario we have outlined
might work.  The point is that we know the $H_u$ mass is driven
negative by renormalization group running
due to the large top quark Yukawa coupling. If the $LH_u$ mass is driven
negative at a sufficiently
high scale, the scenario we have outlined works with no additional nonminimal
Kahler terms present (although as we have emphasized these are there as
counterterms in any case).
Although it might seem dangerous to have the mass run so quickly negative,
since the running also determines the low energy minimum, this is not
necessarily
the case. This is because of the indirect relation between the soft parameters
at high and low temperature, and also because the ``$\mu$" term is not
necessarily
of order $H$, whereas there is certainly a $\mu$ term of order $m_{3/2}$
at  zero temperature. This can allow for our scenario to work where the
parameters
are such as to give the correct low energy minimum. Diego Castano
has verified this scenario numerically.

 To conclude, we have seen that cosmology in a supersymmetric universe is
intriguing and subtle. The existence of flat directions is very important;
so is the fact that there is
 necessarily supersymmetry breaking in the early universe. This means that
 fields are {\it not} frozen; they evolve according to
their classical field equations to their minimum. This
qualitatively changes scenario for coherent field production.
Whether or not  they evolve to the low energy minimum
determines whether or not there is  a Polonyi problem.

Many of our interesting conclusions apply to the
Affleck-Dine Mechanism. We have seen that
 the picture which has emerged is very different
from  the standard scenario in the literature.
We have found a negative mass squared in the early universe
is essential in order to drive the field to a large value (in the
absence of a tuned small mass or a symmetry mechanism).
We have also found a predictive scenario for the field evolution.
The field value varies continuously subsequent to inflation,
tracking the instantaneous minimum. This gives a definite
motivation for the field magnitude at the
  time the AD field is driven to the origin.
Happily,  the $A$ type terms have a magnitude
such that baryon number violation is maximal.
We have found a formula for $n_b/s$ which is roughly
independent of the details of the problem and depends
primarily  on $T_R$ and $n$.
The $LH_u$ direction would be best, because
it does not require an additional source of entropy;
however, almost any direction could work.

\vskip 1in
{\bf Acknowledgements:} I would like to thank Greg Anderson, Diego
Castano, Dan Freedman, Steve Martin, Graham Ross, and Bob Singleton
for useful conversations, and Greg Anderson for his comments on
the manuscript.

\end{document}